# A Frequent Closed Itemsets Lattice-based Approach for Mining Minimal Non-Redundant Association Rules[*]

Bay Vo[1] and Bac Le[2]
1. Faculty of Information Technology, Ho Chi Minh City University of Technology Ho Chi Minh, Vietnam
2. Faculty of Information Technology University of Science, Ho Chi Minh, Vietnam
vdbay@hcmhutech.edu.vn, lhbac@fit.hcmus.edu.vn

*Abstract*

*There are many algorithms developed for improvement the time of mining frequent itemsets (FI) or frequent closed itemsets (FCI). However, the algorithms which deal with the time of generating association rules were not put in deep research. In reality, in case of a database containing many FI/FCI (from ten thousands up to millions), the time of generating association rules is much larger than that of mining FI/FCI. Therefore, this paper presents an application of frequent closed itemsets lattice (FCIL) for mining minimal non-redundant association rules (MNAR) to reduce a lot of time for generating rules. Firstly, we use CHARM-L for building FCIL. After that, based on FCIL, an algorithm for fast generating MNAR will be proposed. Experimental results show that the proposed algorithm is much faster than frequent itemsets lattice-based algorithm in the mining time.*

## 1. Introduction

Mining association rules is divided into two phases: i) Mining FI/FCI and ii) Generating association rules from FI/FCI. There have been many algorithms developed for the phase i) such that Apriori-based [2, 14-15], FP-tree-based [5-7, 16, 23], and IT-tree-based [25-27], etc. However, the algorithms deal with the phase ii) have received little attention. In 1993, Agrawal et al developed a method for mining traditional association rule (TAR) [1]. After that, Apriori algorithm has been proposed [2]. Because TAR contains a lot of redundancies, therefore, minimal non-redundant association rule (MNAR) concept has been proposed [3, 14-15]. The set of MNAR is more compact than TAR in number of generated rules. Besides, the number of FCI is often much smaller than the number of FI, so the time for generating rules from FCI reduces significantly.

Recent years, lattice-based approaches for fast mining association rules have been proposed. In 2009, we proposed an algorithm for mining TAR based on frequent itemsets lattice (FIL) [20]. This work saves a lot of time for generating association rules. Because of based on the lattice, we can determine all child nodes of a given node and need not traverse all FI. After that, a modification of FIL (MFIL) for generating MNAR has been proposed in [22]. MNAR only mines from $X$ to $Y$, where $X$ is a minimal generator, $Y$ is an frequent closed itemset and $X \subset Y$. FIL is modified by adding one field to determine whether a lattice node is a minimal generator or not and one field to determine whether a lattice node is a closed itemset or not. After building the lattice, we can generate MNAR easily.

---

[*] This work was supported by Vietnam's National Foundation for Science and Technology Development (NAFOSTED), project ID: 102.01-2010.02.

The purpose of this paper is to mine MNAR based on frequent closed itemsets lattice and compare it with the algorithm based on MFIL. In section 2, we introduce some basic concepts and related works. Section 3 presents an algorithm for mining MNAR using FCIL. Section 4 discusses our experimental results. Conclusion and future work are in section 5.

## 2. Concepts and Related Works

### 2.1. Transaction Database

Let $I = \{i_1, i_2, \ldots, i_n\}$ be a set of items, $T = \{t_1, t_2, \ldots, t_m\}$ be a set of transaction identifiers (tids or tidset) in a database D. The input database is a binary relation $\delta \subseteq I \times T$. If an item i occurs in a transaction t, we write it as $(i,t) \in \delta$ or $i\delta t$.

Example: Consider database in Table 1

**Table 1. An Example Database**

| TID | Item bought |
|---|---|
| 1 | A, C, T, W |
| 2 | C, D, W |
| 3 | A, C, T, W |
| 4 | A, C, D, W |
| 5 | A, C, D, T, W |
| 6 | C, D, T |

The second transaction can be represented as $\{C\delta 2, D\delta 2, W\delta 2\}$.

### 2.2. Support

Let $D$ be a transaction database and an itemset $X \subseteq I$. The support of $X$, denoted $\sigma(X)$, is number of transactions in $D$ containing $X$.

### 2.3. Frequent Itemset and Frequent Closed Itemset

Itemset $X \subseteq I$ is called to be frequent if $\sigma(X) \geq minSup$ (*minSup* is a minimum support threshold). Let $X$ be a frequent itemset, $X$ is called a frequent closed itemset if there have not any frequent itemset $Y$ such that $X \subset Y$ and $\sigma(X) = \sigma(Y)$.

### 2.4. Minimal Generators [22, 25-26]

Let X be a frequent closed itemset, $X' \neq \varnothing$ is called a *generator* of $X$ if and only if:
i) $X' \subseteq X$ and
ii) $\sigma(X) = \sigma(X')$.

Let $G(X)$ denote the set of *generator* of X. We say that $X' \in G(X)$ is a *minimal generator* if it has no subset in $G(X)$. Let $mGs(X)$ denote the set of all *minimal* *generators* of X. By definition, $mGs(X) \neq \varnothing$ since if there is no proper *generator* then $X$ is a *mG* of X.

### 2.5. Mining FCI

Mining FCI is divided into four categories [9, 24]:
i) Test-and-generate (Close [15], A-Close [14]): Using level-wise approach to discover FCI. All of them are based on the Apriori algorithm.
ii) Divide-and-conquer (Closet [16], Closet+ [23], FPClose [6]): using compact data structure (extended from FP-tree) to mine FCI.
iii) Hybrid (CHARM [27], CloseMiner [18]): using both test-and-generate and divide-and-conquer to mine FCI. They are based on vertical data format to transform the database into item – tidlist and develop properties to prune fast non-closed itemsets.
iv) Hybrid without duplication (DCI-Close [12], LCM [19], PGMiner [13]): they differ from hybrid in that they do not use "subsume checking". Therefore, they do not need storage of FCI in main memory and need not use hash tables as CHARM.

### 2.6. Mining MNAR/NAR

Mining MNAR was proposed in 1999 by Pasquier et al. [14-15]. Firstly, the authors mined all FCI by computing closure of minimal generators. After that, they mined all MNAR by generating rules with confidence = 100% from $mGs(X)$ to X ( X is a frequent closed itemset) and generating rules with the confidence < 100% from $mGs(X)$ to Y (X, Y are frequent closed itemsets and $X \subset Y$). In 2000, Zaki proposed the method to mine NARs [25]. He was based on FCI and theirs *mGs* to mine NARs. This approach only mined the rules that their left hand side and right hand side are minimal in the set of rules that have the same support and confidence. In 2004, Zaki published his paper with some extensions [26].

### 2.7. Building Frequent (closed) Itemsets Lattice

Zaki and Hsiao proposed CHARM-L [27], which is an extension of CHARM, for building a frequent closed itemset lattice. We presented an extension of the Eclat algorithm [27] for building a frequent itemset lattice (FIL) [20]. A modification of the frequent itemset lattice for mining MNAR was also presented in [22].

In this paper, we extend the lattice-based approach for quickly mining MNAR. Firstly, CHARM-L algorithm for building FCIL will be applied. After that, based on FCIL, a mining approach for MNAR based on the obtained FCIL is designed.

### 2.8. Generating Association Rules from Frequent Itemsets Lattice

In [20], we have proposed an algorithm for mining traditional association rules from FIL. This algorithm uses the relation between two nodes in lattice for fast traversing all child nodes of a given node. This approach is more efficient than directly mining from frequent itemsets (using hash table) [7]. A modification of frequent itemsets lattice (MFIL) for mining MNAR was proposed in [22].

## 3. Generating Minimal Non-redundant Association Rules from FCIL

**Definition 3.1. General rule [22]**

Let two rules $R_1$: $X_1 \rightarrow Y_1$ and $R_2$: $X_2 \rightarrow Y_2$, $R_1$ is said more general than $R_2$, denoted $R_1 \propto R_2$, if and only if $X_1 \subseteq X_2$ and $Y_2 \subseteq Y_1$.

**Definition 3.2. Redundant rule [22]**

Let $R = \{R_1, R_2, \ldots, R_n\}$ be the set of rules which have the same support and confidence. Rule $R_j$ is redundant if in $R$ exists the rule $R_i$ such that $R_i \propto R_j$ ($i \neq j$).

**Theorem 3.1 [22].** MNAR with the confidence = 100% are only generated from $X' \rightarrow X$ ($\forall X' \in mGs(X)$, $X$ is a FCI).

**Theorem 3.2 [22].** MNAR with the confidence < 100% are only generated from $X' \rightarrow Y$ ($\forall X' \in mGs(X)$, $X, Y \in$ FCI, and $X \subset Y$).

Based on the theorem 3.1 and theorem 3.2, MNAR are only mined from $X$ to $Y$, where $X$ is a minimal generator, $Y$ is a frequent closed itemset and $X \subset Y$. Therefore, we modify CHARM-L [27] to build FCIL and mine all minimal generators of FCI by using MG-CHARM [21].

### 3.1. Algorithm for Generating MNAR from FCIL

```
Input: Frequent closed itemsets lattice with the root node L_r and minConf
Output: MNAR satisfy minConf
Method:
GENERATING_MNAR_FROM_FCIL(L_r)
1.   MNAR = ∅
2.   for each L_c ∈ L_r.Children do
3.      EXTEND_MNAR_FCIL( L_c )
4.   return MNAR
EXTEND_MNAR_FCIL(L_c)
5.   if L_c.flag = False then
6.      GENERATE_RULE(L_c)
7.      L_c.flag = True
8.   for each L_s ∈ L_c.Children do
9.      EXTEND_MNAR_FCIL( L_s )
GENERATE_RULE(L_c)
10.  Queue = ∅
11.  FIND_RULE(L_c, L_c, 1.0)
12.  for all L_s ∈ L_c.Children do
13.     Add L_s to Queue
14.     Mark L_s
15.  while Queue ≠ ∅ do
16.     L = Get an element from Queue
17.     conf = L.sup/ L_c.sup
18.     if conf ≥ minConf then
19.        FIND_RULE(L_c, L_s, conf)
20.        for all L_s ∈ L.Children do
21.           if L_s is not marked then
22.              Add L_s to Queue
23.              Mark L_s
FIND_RULE(L_c, L_s, conf)
24.  for all Z ∈ L_c.mG do
25.     if L_s.itemset \ Z ≠ ∅ then
26.        MNAR = MNAR ∪ { Z → L_s.itemset \ Z (L_s.sup,conf)}
```

**Figure 1. Algorithm for mining minimal non-redundant association rules from FCIL**

Firstly, the algorithm traverses all child nodes $L_c$ of the root node $L_r$, then it calls EXTEND_MNAR_FCIL($L_c$) function (line 3) to traverse all nodes in the FCIL (recursively and marks the visited nodes by turn flag on, lines 5-9). Considering GENERATE_RULE($L_c$) function, this function used a queue to traverse all child nodes of $L_c$ (and marking all visited nodes to reject coincide). For each child node $L$ of $L_c$, we compute the confidence of rules that will generate from $L_c$ to $L$, if the confidence satisfies *minConf* then function FIND_RULE is called (lines 17-19) to generate all rules from minimal generators of $L_c$.itemset to $L$.itemset (lines 24-26).

### 3.2. Illustration

Using CHARM-L [27] and MG-CHARM [21], we have the frequent closed itemsets lattice of the database in Table 1 with *minSup* =50% as follow:

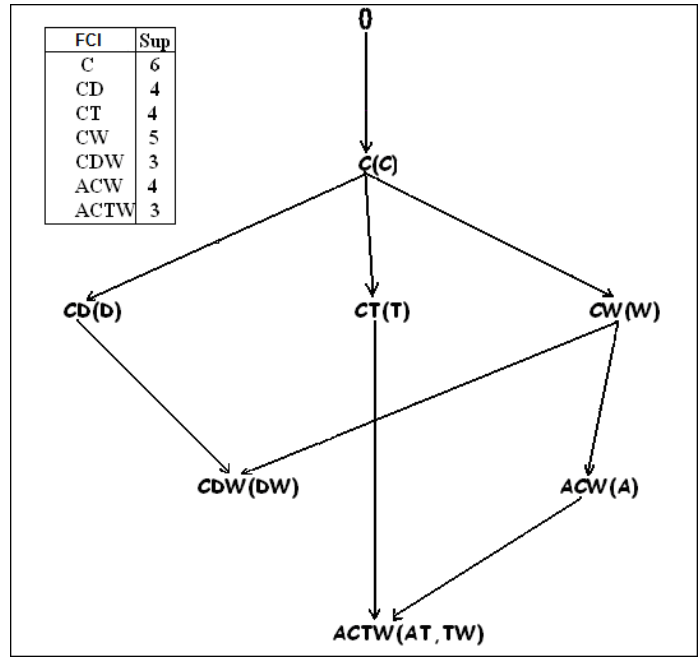

**Figure 2. Frequent closed itemsets lattice and their minimal generators of the database in Table 1.**

Consider the process of generating MNAR with the *minConf* = 80% from node {CW} of the lattice (in Figure 2), we have:
- At first, Queue = ∅. After that, the algorithm will call FIND_RULE ({CW}, {CW}, 1.0). This function will generate rule from *mG* of CW (ie., W) to CW with the confidence is 100%, we have rule $W \xrightarrow{5,1.0} C$.
- The child nodes of {CW} are {{CDW}, {ACW}}, so they are added to Queue ⇒ Queue = {{CDW}, {ACW}}.
- Because Queue ≠ ∅:
  - $L$ = {CDW} (Queue = {{ACW}}).
  - The confidence of rules from CW to CDW is 3/5 < minConf, the algorithm will not call function FIND_RULE.
  - Next, because Queue ≠ ∅:

- $L$ = {ACW} (Queue = $\varnothing$).
- The confidence of rules from CW to ACW is 4/5 ≥ minConf, the algorithm will call function FIND_RULE ({CW}, {ACW}, 4/5). This function will generate rule $W \xrightarrow{4,4/5} AC$. After that, algorithm will add all child nodes of ACW to Queue $\Rightarrow$ Queue = {{ACTW}}
- Next, because Queue ≠ $\varnothing$:
  - $L$ = {ACTW} (Queue = $\varnothing$).
  - The confidence of rules from CW to ACTW is 3/5 < minConf, the algorithm will not call the function FIND_RULE.
  - Next, because Queue = $\varnothing$, stop.

## 4. Experimental Results

All experimental results described below have been performed on a Centrino core 2 duo (2×2.53 GHz), 4GBs RAM memory, Windows 7. Algorithms were coded in C# (2008). The experimental databases from http://fimi.cs.helsinki.fi/data/ (downloaded on April 2005) were downloaded to perform the test with theirs features displayed in Table 2.

**Table 2. Features of Databases**

| Databases | #Trans | #Items |
|---|---|---|
| Chess | 3196 | 76 |
| Mushroom | 8124 | 120 |
| Pumsb | 49046 | 7117 |
| Connect | 67557 | 130 |
| Accidents | 340183 | 468 |

Figures from 3 to 7 present the executtion time of two algorithms for mining MNAR based on MFIL and FCIL.

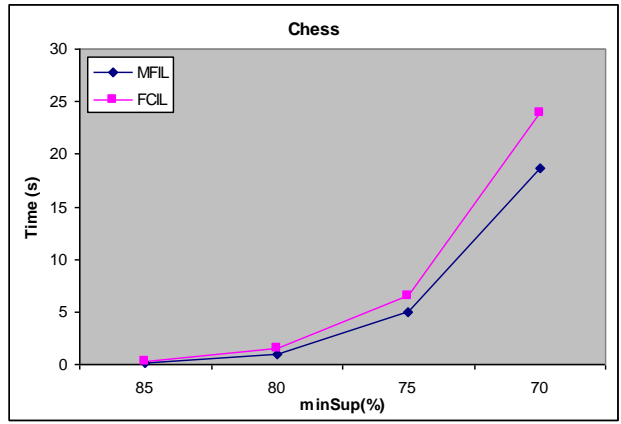

**Figure 3. Execution time of the two algorithms for Chess under different minSup values**

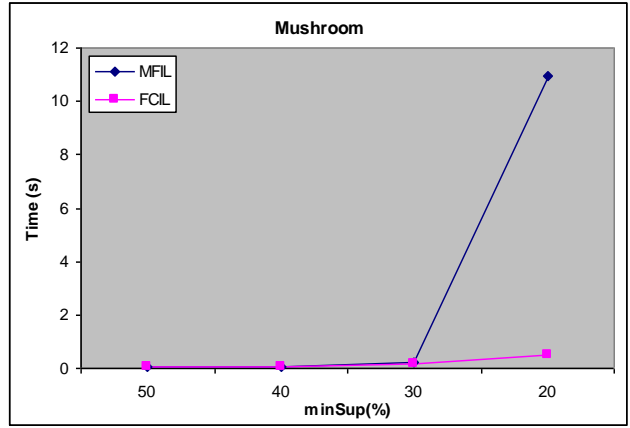

**Figure 4. Execution time of the two algorithms for Mushroom under different minSup values**

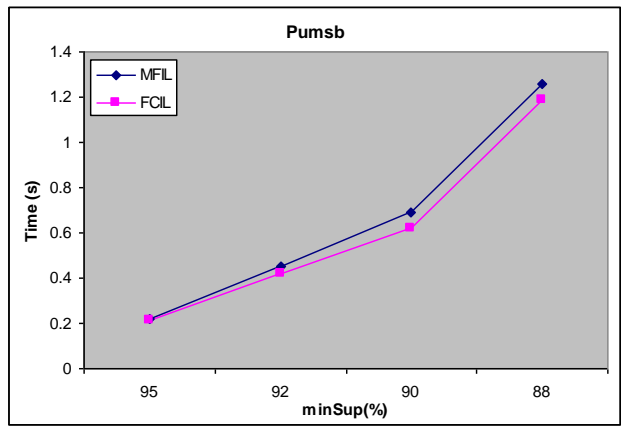

**Figure 5. Execution time of the two algorithms for Pumsb under different minSup values**

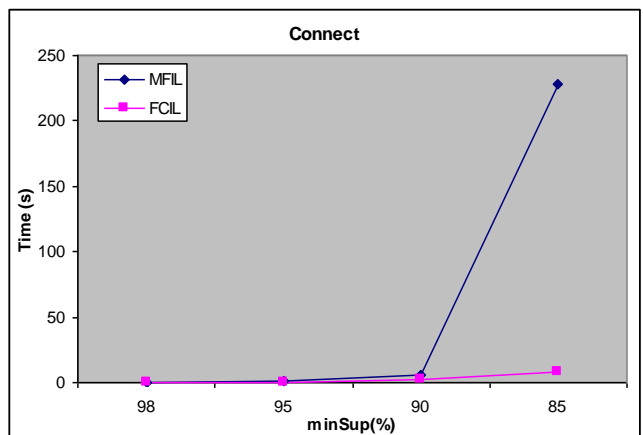

**Figure 6. Execution time of the two algorithms for Connect under different minSup values**

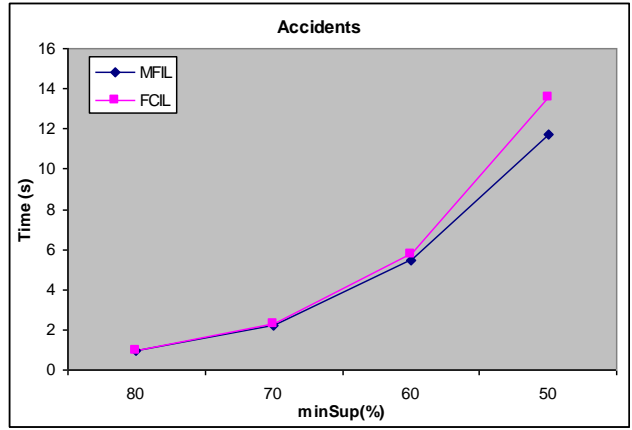

**Figure 7. Execution time of the two algorithms for Accidents under different minSup values**

Results from figures 3-7 show that the execution time of mining MNAR based on FCIL is faster than that of based on MFIL [22]. Especially, when the number of FI in MFIL is much larger than the number of FCI in FCIL. For example, consider Mushroom with minSup = 20%, the number of FI is 53583 while the number of FCI is only 1200.

If we only measure the time of generated rules (without considering the time of mining FCI and building lattice), the results for the five databases are shown from Figure 8 to Figure 12.

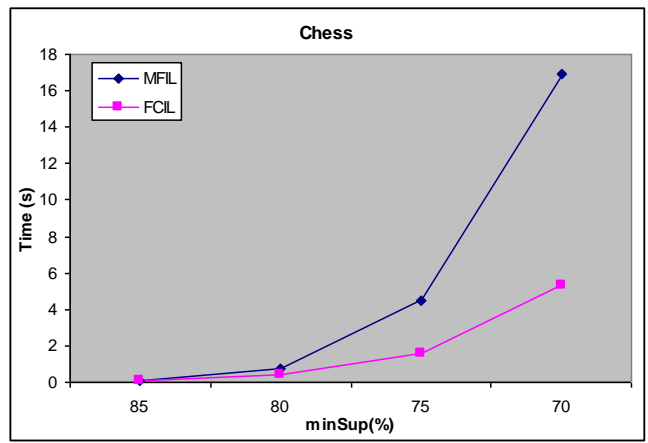

**Figure 8. Execution time of the two algorithms for Chess under different minSup values (without mining FCI and building lattice)**

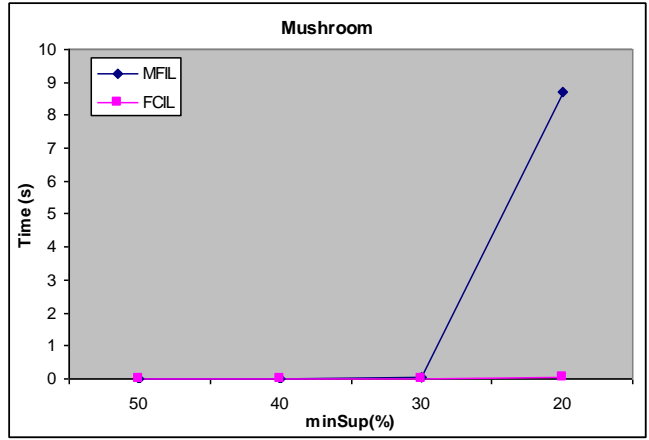

**Figure 9. Execution time of the two algorithms for Mushroom under different minSup values (without mining FCI and building lattice)**

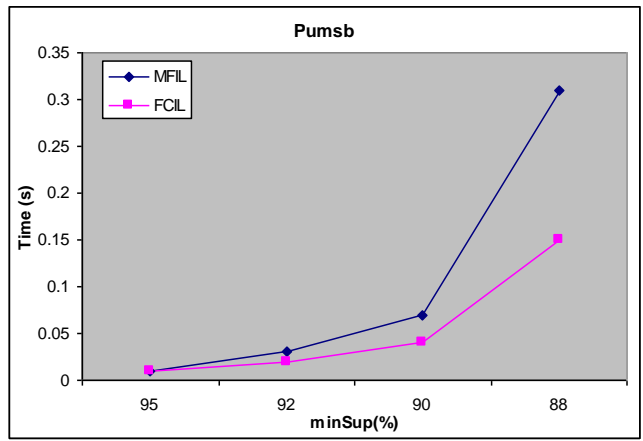

**Figure 10. Execution time of the two algorithms for Pumsb under different minSup values (without mining FCI and building lattice)**

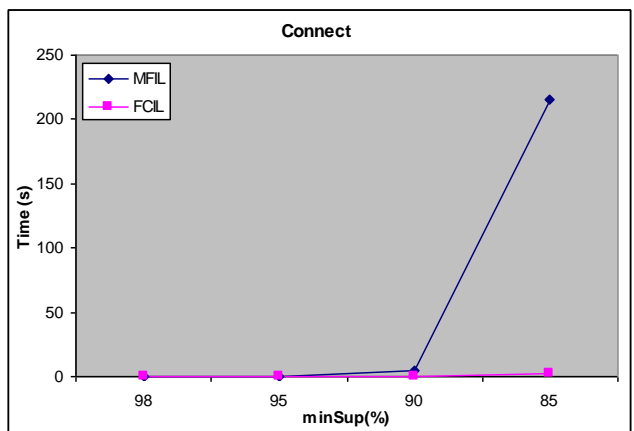

**Figure 11. Execution time of the two algorithms for Connect under different minSup values (without mining FCI and building lattice)**

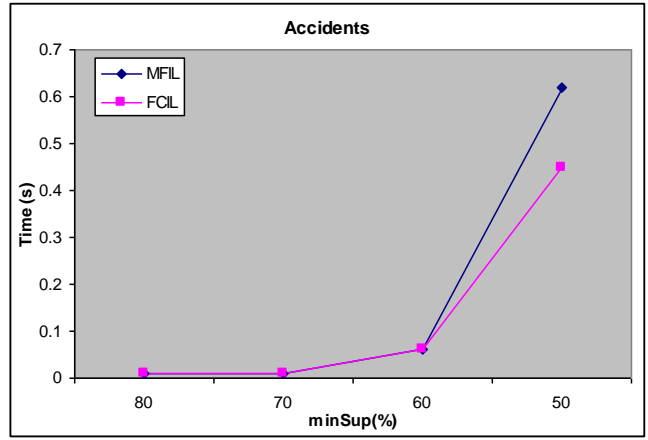

**Figure 12. Execution time of the two algorithms for Accidents under different minSup values (without mining FCI and building lattice)**

Results from Figure 8 to Figure 12 show that generating MNAR based on FCIL is always faster than that of based on MFIL in all databases if we do not consider the time of mining FCI and building lattice.

## 5. Conclusion and Future Work

This paper has proposed a method for mining minimal non-redundant association rules based on frequent closed itemsets lattice. Experimental results show that the proposed algorithm is more efficient than that of mining MNAR from a modification frequent itemsets lattice. Although building lattice consumes a bit of time for updating parent-child relationship between nodes and memory for the storage these relations, but the algorithm saves a lot of time in generating rules.

In future, we will study how to build FCIL faster. Besides, a method for mining efficient association rules will be discussed.

## References

[1] R. Agrawal, T. Imielinski, A. Swami, "Mining Association Rules between Sets of Items in Large Databases", *Proceedings of the 1993 ACM SIGMOD Conference*, Washington DC, USA, May 1993, pp. 207 – 216.
[2] R. Agrawal, R. Srikant, R., "Fast Algorithms for Mining Association Rules", VLDB'94, pp. 487 – 499.
[3] Bastide, Y., Pasquier, N., Taouil, R., Stumme, G., Lakhal, L., "Mining Minimal Non-Redundant Association Rules using Frequent Closed Itemsets", 1st International Conference on Computational Logic, 2000, pp. 972 – 986.
[4] V. Choi, "Faster Algorithms for Constructing a Concept (Galois) Lattice", arXiv:cs.DM/0602069, 2006.
[5] G. Grahne, J. Zhu, "Efficiently Using Prefix-trees in Mining Frequent Itemsets", FIMI'03 Workshop on Frequent Itemset Mining Implementations (held with IEEE ICDM'03), 2003, pp. 123 – 132.
[6] G. Grahne, J., Zhu, "Fast Algorithms for Frequent Itemset Mining Using FP-Trees", IEEE Trans. Knowl. Data Eng. 17(10), 2005, pp. 1347-1362.
[7] J. Han, M. Kamber, "Data Mining: Concept and Techniques", Second Edition, Morgan Kaufmann Publishers, Sansome Street, San Francisco, CA, USA, 2006, pp. 239-241.
[8] http://fimi.cs.helsinki.fi/data/ (downloaded on April 2005)
[9] D.G. Kourie, S. Obiedkov, B. W. Watson, D. V. D., Merwe, "An Incremental Algorithm to Construct a Lattice of Set Intersections", Science of Computer Programming 74, 2009, pp. 128 – 142.
[10] S. O. Kuznetsov, S. A. Obiedkov, "Comparing Performance of Algorithms for Generating Concept Lattices", J. Exp. Theor. Artif. Intell. 14(2-3), 2002, pp. 189-216.

[11] A. J. T. Lee, C. S. Wang, W. Y. Weng, J. A. Chen, H. W. Wu, "An Efficient Algorithm for Mining Closed Inter-transaction Itemsets", Data & Knowledge Engineering (66), 2008, pp. 68 – 91.
[12] B. Lucchese, S. Orlando, R. Perego, "Fast and Memory Efficient Mining of Frequent Closed Itemsets", ", IEEE Transactions on Knowledge and Data Engineering, Vol. 18, No. 1, 2006, pp. 21 – 36.
[13] H. D. K. Moonestinghe, S. Fodeh, P. N. Tan, "Frequent Closed Itemsets Mining using Prefix Graphs with an Efficient Flow-based Pruning Strategy", Proceedings of 6th ICDM, Hong Kong, 2006, pp. 426 – 435.
[14] N. Pasquier, Y. Bastide, R. Taouil, L. Lakhal, "Discovering Frequent Closed Itemsets for Association Rules", Proc. Of the 5th International Conference on Database Theory, LNCS, Springer-Verlag, Jerusalem, Israel, 1999, pp. 398 – 416.
[15] N. Pasquier, Y. Bastide, R. Taouil, L. Lakhal, "Efficient Mining of Association Rules using Closed Itemset Lattices", Information Systems 24 (1), 1999, pp. 25 – 46.
[16] J. Pei, J. Han, R. Mao, "CLOSET: An Efficient Algorithm for Mining Frequent Closed Itemsets", Proc. of the 5th ACM-SIGMOD Workshop on Research Issues in Data Mining and Knowledge Discovery, Dallas, Texas, USA, 2000, pp. 11 – 20.
[17] Y. Shuqun, D. Shuliang, C. Shengzen, D. Qiulin, "An Algorithm of Constructing Concept Lattices for CAT with Cognitive", Knowledge-based Systems 21, 2008, pp. 852 – 855.
[18] N. G. Singh, S. R. Singh, A. K. Mahanta, "CloseMiner: Discovering Frequent Closed Itemsets using Frequent Closed Tidsets", Proc. of the 5th ICDM, Washington DC, USA, 2005, pp. 633 – 636.
[19] T. Uno, T. Asai, Y. Uchida, H. Arimura, "An Efficient Algorithm for Enumerating Closed Patterns in Transaction Databases", Proc. of the 7th International Conference on Discovery Science, LNCS, Springer Verlag, Padova, Italy, 2004, pp. 16 – 31.
[20] B. Vo, B. Le, "Mining traditional association rules using frequent itemsets lattice", 39th International Conference on Computers & Industrial Engineering, July 6 – 8, Troyes, France, 2009, pp. 1401 – 1406.
[21] B. Vo, B. Le, "Fast algorithm for mining minimal generators of frequent closed itemsets and their applications", the 39th International Conference on Computers & Industrial Engineering, July 6 – 8, Troyes, France, IEEE, 2009, pp. 1407 – 1411.
[22] B. Vo, B. Le, "Mining Minimal Non-redundant Association Rules using Frequent Itemsets Lattice", Int. J. Intelligent Systems Technologies and Applications, Vol. 10, No. 1, 2011, pp.92–106.
[23] J. Wang, J. Han, J. Pei, "CLOSET+: Searching for the Best Strategies for Mining Frequent Closed Itemsets", ACM SIGKDD International Conference on Knowledge Discovery and Data Mining, 2003, pp. 236 – 245.
[24] S. B. Yahia, T. Hamrouni, E. M., Nguifo, "Frequent Closed Itemset based Algorithms: A thorough Structural and Analytical Survey", ACM SIGKDD Explorations Newsletter 8 (1), 2006, pp. 93 – 104.
[25] M. J. Zaki, "Generating Non-Redundant Association Rules", Proc. of the 6th ACM SIGKDD International Conference on Knowledge Discovery and Data Mining, Boston, Massachusetts, USA, 2000, pp. 34 – 43.
[26] M. J. Zaki, "Mining Non-redundant Association Rules", Data Mining and Knowledge Discovery, Kluwer Academic Publishers, Hingham, MA, USA, 2004, pp. 223–248.
[27] M. J. Zaki, C. J., Hsiao, "Efficient Algorithms for Mining Closed Itemsets and Their Lattice Structure", IEEE Transactions on Knowledge and Data Engineering, Vol. 17, No 4, April 2005, pp. 462-478.

## Authors

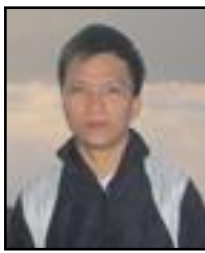

**Bay Vo** is currently Ph.D student at Computer Science department, University of Science, Ho Chi Minh City, Vietnam. He received his Bachelor of Science (2002) and Master of Science (2005) degrees from University of Science, Ho Chi Minh City, Vietnam. His research interests include association rules, classification, data mining in multidimensional database, distributed database and privacy preserving in data mining.

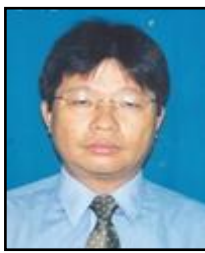

**Bac Le** received the BSc degree, in 1984, the MSc degree, in 1990, and the PhD degree in Computer Science, in 1999. He is an Associate Professor, Vice Dean of Faculty of Information Technology, Head of Department of Computer Science, University of Science, Ho Chi Minh City. His research interests are in Artificial Intelligent, Soft Computing, and Knowledge Discovery and Data Mining.